\documentclass{ifacconf}

\usepackage{graphicx}      % include this line if your document contains figures
\usepackage{natbib}        % required for bibliography

\usepackage{bm}
\usepackage{bbm}
\usepackage{multirow}
\usepackage{amsmath}
\DeclareMathOperator*{\argmax}{arg\,max}

\DeclareMathOperator*{\diag}{diag}
\DeclareMathOperator*{\col}{col}
\DeclareMathOperator*{\row}{row}

\DeclareMathAlphabet{\mathcal}{OMS}{cmsy}{m}{n}
\usepackage{color}

\usepackage{mathbbol}

\newtheorem{assumption}{Assumption}

\newtheorem{remark}{Remark}

\newtheorem{problem}{Problem}
\usepackage{bbm}
 
\DeclareMathOperator{\Tr}{Tr}

\usepackage{algorithm}
\usepackage{algpseudocode}
\usepackage{url}

\begin{document}
\begin{frontmatter}

\title{$\mathcal{H}_2$-norm transmission switching to improve synchronism of low-inertia power grids} 

\thanks[footnoteinfo]{This work was supported by the Research Grants Council of the Hong Kong Special Administrative Region under under the General Research Fund (GRF) through Project No. 17209419.}

\author[First]{Tong Han}
\author[First]{David J.Hill}

\address[First]{The University of Hong Kong, 
   Hong Kong,\\ (e-mail: hantong, dhill@eee.hku.hk).}
% \address[Second]{Colorado State University, 
%    Fort Collins, CO 80523 USA (e-mail: author@lamar. colostate.edu)}

\begin{abstract}          
    This paper investigates the utilization of transmission switching to improve synchronization performance of low-inertia grids. The synchronization performance of power girds is first measured by the $\mathcal{H}_2$ norm of linearized power systems. Laplacian-based bounds and a close-form formulation of the $\mathcal{H}_2$-norm synchronization performance metric are derived to reveal the influence of network structure on synchronization performance. Furthermore, a transmission switching approach is developed by analyzing the sensitivity of the $\mathcal{H}_2$-norm metric to perturbation of network susceptance. Effectiveness of the proposed approach to improve synchronization performance is demonstrated using the SciGRID network for Germany.
\end{abstract}


\end{frontmatter}
%===============================================================================

\section{Introduction}

Power grids are evolving toward 100\% renewable energy to stem catastrophic climate change. Conventional synchronous generators, whose inertia and damping are essential for synchronism of power grids, will be steadily substituted by inverter-interfaced generation. This transition, however, can cause new challenges for power grid operations, one of which is reduction of synchronization performance due to the resulting low, time-varying and heterogeneous system inertia \citep{4-999}.

To tackle the deterioration of synchronism, much effort has been made to design control strategies of inverters and optimize parameters of inverter's control loop systematically. These efforts essentially improve synchronization performance by regulating node (i.e., inverter) dynamic properties of power grids. However, from a perspective of dynamic networks, not only node properties but also network structures can influence network's dynamic behaviors. For conventional or inverter-interfaced power grids, the key role of grid topology for synchronization and some stability issues has already been revealed \citep{4-480, 4-234, 4-10}. Therefore, it is of great potential to leverage the flexibility in the topology to tackle the new challenge arising from the transition of power grids. 

In transmission networks, transmission switching has been demonstrated to be effective for reducing dispatch cost \citep{4-62}, relieving overload and voltage violation \citep{4-70, 4-71}, enhancing the small-signal stability margin \citep{4-361}, etc \citep{4-581, 4-7}. In contrast to regulation of inverters' dynamic properties which requires support of energy storage devices, transmission switching, in the physical layer, only relies on breakers and communication networks which are generally fully equipped for modern power grids. Considering the high cost of large-scale energy storage, transmission switching can be one of techniques for bringing a economical package to solve synchronism and frequency stability problems of low-inertia power grids. It should be pointed out that transmission switching is unable to solve these problems individually and energy storage-based frequency regulation is still indispensable.

% or optimal transmission switching (OTS) when optimal switching being pursued

In this work, we utilize transmission switching as a mean to improve synchronization performance for low-inertia power grids. Contributions of this paper are twofold:
\begin{itemize}
    \item {We investigate the impact of network structure on synchronization performance measured by the $\mathcal{H}_2$ norm of linearized power systems, where Laplacian-based bounds and a close-form formulation of the $\mathcal{H}_2$ norm are derived.}
    \item {A transmission switching approach is developed to improve synchronization performance by analyzing sensitivity of the $\mathcal{H}_2$ norm to perturbation of network susceptance.}
\end{itemize}

\section{Dynamic Models}
Consider a lossless transmission power grid denoted by an undirected graph $\mathcal{G}(\mathcal{V}, \mathcal{E})$ where $\mathcal{V}$ is the set of vertices (buses) and $\mathcal{E}$ is the set of edges (branches). The power grid is augmented with the internal buses of synchronous generators. $\mathcal{V}$ consists of $\mathcal{V}_{\rm{L}}$, $\mathcal{V}_{\rm{S}}$ and $\mathcal{V}_{\rm{FM}}$, denoting sets of load buses and buses with neither load nor power generation, synchronous generator buses and grid-forming inverter buses, respectively. $\mathcal{V}_{\rm{SF}} = \mathcal{V}_{\rm{S}} \cup \mathcal{V}_{\rm{FM}}$. $\underline{\mathcal{V}}_{\rm{SF}}$ equals to $\mathcal{V}_{\rm{SF}}$ with the first bus deleted. $\underline{\mathcal{V}} = \underline{\mathcal{V}}_{\rm{SF}} \cup {\mathcal{V}}_{\rm{L}} $. $\mathcal{E}$ can be divided into $\mathcal{E}_{\rm{SF}} \subseteq \mathcal{V}_{\rm{SF}} \times \mathcal{V}_{\rm{L}}$ and $\mathcal{E}_{\rm{L}} \subseteq \mathcal{V}_{\rm{L}} \times \mathcal{V}_{\rm{L}}$. The induced graph of $\mathcal{G}$ by $\mathcal{V}_{\rm{L}}$, denoted as $\widetilde{G}(\mathcal{V}_{\rm{L}}, \mathcal{E}_{\rm{L}})$ contains branches which can be switched. Denote by $V_i$, $\theta_i$ and $\omega_i$ the voltage magnitude, voltage phase angle and angular frequency of bus $i$, respectively. Additionally, denote by $\bm{L}(\mathcal{G}, \bm{W})$ the Laplacian matrix of weighted graph $\mathcal{G}$ with $\bm{W}$ being the diagonal weighted matrix, and $\underline{\bm{L}}(\mathcal{G}, \bm{W})$ called the reduced Laplacian matrix, the principal matrix of $\bm{L}(\mathcal{G}, \bm{W})$ formed by deleting the first row and column. $\bm{E}_{\mathcal{G}}$ and $\bm{E}_{\tilde{\mathcal{G}}}$ are incidence matrices of $\mathcal{G}$ and $\widetilde{\mathcal{G}}$, respectively. $\underline{\bm{E}}_{\mathcal{G}}$ is formed by deleting the first row of $\bm{E}_{\mathcal{G}}$. 

The structure preserving property is vital to dynamic models employed for the transmission switching problem. Hence following the previous work by \cite{4-999-102} and \cite{4-10}, we adopt the frequency-dependent load model and exert singular perturbation to buses with neither load nor power generation. Then the dynamic of load buses is given by
\begin{equation}\label{eq-model-1}
    d_i \dot{\theta_i}  = p_{{\rm{in}},i} -  \sum_{j \in \mathcal{V}} V_i V_j b_{ij} \sin(\theta_i - \theta_j)    ~~~~ \forall i \in \mathcal{V}_{\rm{L}}
\end{equation}
where $d_i$ and $p_{{\rm{in}},i}$ are the frequency coefficient and opposite of load power of bus $i$, respectively. Buses with zero power injection are regarding as load buses with $d_i$ being singularly perturb as $d_i = \epsilon$ where $\epsilon$ is a sufficiently small positive number. $b_{ij}$ is the susceptance between buses $(i,j) \in \mathcal{E}$. The voltage magnitude $V_i$ of each bus is assumed to be constant.

Dynamics of synchronous generator and grid-forming inverter buses are given by
\begin{equation}\label{eq-model-2}
    \begin{aligned}
        & \dot{\theta_i} = \omega_i ~~~ \forall i \in \mathcal{V}_{\rm{SF}}\\
        & m_i \dot{\omega}_i \!=\! - d_i \omega \!+\!  p_{{\rm{in}},i} \!-\! \sum_{j \in \mathcal{V}} \!\! V_i V_j b_{ij} \sin(\theta_i \!-\! \theta_j) ~~~ \forall i \in \mathcal{V}_{\rm{SF}}
    \end{aligned}
\end{equation}
where $m_i$ and $d_i$ are the inertia (or virtual inertia) and damping coefficients of synchronous generators or grid-forming inverters, respectively; $p_{{\rm{in}},i}$ is the set point of active power generation.
 
To obtain the state-space model of power grids, we take the first bus in $\mathcal{V}_{\rm{SF}}$ as the angle reference, and a new vector is introduced as
\begin{equation}\label{eq-model-7}
    \bm{\alpha} 
    =  \col(\bm{\alpha}_{\rm{SF}}, \bm{\alpha}_{\rm{L}}) 
     = \bm{T} \bm{\theta} \in \mathbb{R}^{|\underline{\mathcal{V}}|} 
\end{equation}
with $\bm{T} = \row(-  \bm{1}_{|\underline{\mathcal{V}}|}, \bm{I}_{|\underline{\mathcal{V}}| }) \in \mathbb{R}^{|\underline{\mathcal{V}}|\times |\mathcal{V}|} $ being the transformation matrix \citep{4-10}. $\bm{T}_{\rm{SF}} \in \mathbb{R}^{  |\underline{\mathcal{V}}| \times |\underline{\mathcal{V}}_{\rm{SF}}| }$ and $\bm{T}_{\rm{L}}  \in \mathbb{R}^{|\underline{\mathcal{V}}| \times |\underline{\mathcal{V}}_{\rm{L}}| }$ consist of columns of $\bm{T}$ corresponding to buses $\underline{\mathcal{V}}_{\rm{SF}}$ and buses $\mathcal{V}_{\rm{L}}$, respectively. 

Now we consider a disturbance input vector $\Delta \bm{u}$ satisfying $\bm{\Lambda}^{\frac{1}{2}} \Delta \bm{u} = \underline{\bm{p}}_{\rm{in}} - \underline{\bm{p}}_{\rm{in}}^0$. $\Delta \bm{u} = \col( \Delta \bm{u}_i), \forall i \in \underline{\mathcal{V}}$ indicates the type of power disturbances, and $\bm{\Lambda} = \diag({\Lambda}_i), \forall i \in \underline{\mathcal{V}}$ is the parametric matrix to model the location and strength of power disturbances. Additionally, denote by $\Delta \bm{y}$ the performance output vector of linearized power grids. With state variables being $\bm{x} = \col(\bm{\alpha}, \underline{\bm{\omega}}_{\rm{SF}} )$, the state-space model of linearization of power grids around the equilibrium point $\bm{x}^{0} = \col(\bm{\alpha}^{0}, \underline{\bm{\omega}}_{\rm{SF}}^{0} )$ is given as
\begin{equation}\label{eq-model-18}
    \begin{bmatrix}
        \Delta \dot{\bm{x}} \\
        \Delta \bm{y}
    \end{bmatrix}
    =
    \begin{bmatrix}
        \bm{A} & \bm{B} \\
        \bm{C} & \bm{O}
    \end{bmatrix}
    \begin{bmatrix}
        \Delta \bm{x} \\
        \Delta \bm{u}
    \end{bmatrix}
\end{equation}
with
\begin{equation}\label{eq-bound-1}
    \begin{aligned}
        \bm{A}  = 
         \begin{bmatrix}
            - \bm{T}_{\rm{L}} \bm{D}_{\rm{L}}^{-1} \bm{T}_{\rm{L}}^T \underline{\bm{L}}(\mathcal{G}, \bm{W}_{\rm{p}}) & \bm{T}_{\rm{SF}} \\
            - \underline{\bm{M}}_{\rm{SF}}^{-1} \bm{T}_{\rm{SF}}^T \underline{\bm{L}}(\mathcal{G}, \bm{W}_{\rm{p}}) & - \underline{\bm{M}}_{\rm{SF}}^{-1} \underline{\bm{D}}_{\rm{SF}}
        \end{bmatrix}
    \end{aligned}
\end{equation}

\begin{equation}\label{eq-bound-2}
    \bm{B} = 
    \begin{bmatrix}
        O & \bm{T}_{\rm{L}} \bm{D}_{\rm{L}}^{-1} \bm{\Lambda}_{\rm{L}}^{\frac{1}{2}}\\
        \underline{\bm{M}}_{\rm{SF}}^{-1} \bm{\Lambda}_{\rm{SF}}^{\frac{1}{2}} & O 
    \end{bmatrix}
\end{equation}
where $\bm{D}_{\rm{L}} = \diag(d_i), \forall i \in \mathcal{V}_{\rm{L}}$, $\underline{\bm{D}}_{\rm{SF}} = \diag(d_i), \forall i \in \underline{\mathcal{V}}_{\rm{SF}}$, $\underline{\bm{M}}_{\rm{SF}} \!=\! \diag(m_i), \forall i \!\in\! \underline{\mathcal{V}}_{\rm{SF}}$ and $\bm{W}_{\rm{p}} \!=\! \bm{B}_{\rm{V}} \frac{\partial \sin(\underline{\bm{E}}_{\mathcal{G}}^{T}  \bm{\alpha}^0) }{\partial (\underline{\bm{E}}_{\mathcal{G}}^{T} \bm{\alpha}^0)  } $ with $\bm{B}_{\rm{V}} = \diag(V_i V_j b_{ij}), \forall (i,j) \in \mathcal{E}$. Denote by $\bm{G}(s)$ the transfer matrix between the disturbance input $\Delta \bm{u}$ and the performance output $\Delta \bm{y}$.

\section{Synchronization performance metrics}\label{section-3}

Synchronization of power grids is generally understood as an integration of phase cohesiveness and frequency synchronization (or frequency boundedness) \citep{4-999-172, 4-999-174}. While the extreme of angle difference and frequency determines whether the system remains synchronous, overall performance metrics to evaluate synchronism are preferred for optimization problems \citep{4-481, 4-999-208}. Here we define the following metric $\mathcal{S}$ to evaluate the synchronization performance of power grids regarding a given disturbance with its time-domain response: 
\begin{equation}\label{eq-model-26}
    \begin{aligned}
        & \mathcal{S}(T_f) =  \\
        &   \frac{\xi(T_f, \mathcal{s})}{T_f} \!\!\! \int_{0}^{T_f} \!\!\! \big [ \!\!
        \sum_{(i, j) \in \mathcal{E} } \!\!\! \bm{W}_{1, ij} (\Delta \theta_i - \Delta \theta_j)^2
        \!+\!\! 
        \sum_{i \in \underline{\mathcal{V}}_{\rm{SF}}} \!\! \bm{W}_{2, i} \Delta {\omega_i}^2
        \big ] \text{d}t
    \end{aligned}
\end{equation}
where $[0, T_f]$ is the time horizon of interest; function $\xi(T_f, \mathcal{s})$ returns $T_f$ when $T_f = +\infty$ and the integral term denoted as $\mathcal{s}$ is finite, and $1$ otherwise; $\bm{W}_{1, ij}$ and $\bm{W}_{2, ij}$ are positive weighting factors or scalars. Denote by $\bm{W}_1$ and $\bm{W}_2$ matrices $\diag(\bm{W}_{1, ij})$ and $\diag(\bm{W}_{2, ij})$, respectively. $\mathcal{S}(T_f)$ expresses an average synchronization performance in time horizon $[0, T_f]$ except in the case where we consider a infinite time horizon but $\mathcal{s}$ is finite. In such case, $\mathcal{S}(T_f) = \mathcal{s}$ should be understood as an accumulative synchronization performance in time horizon $[0, T_f]$. Furthermore, with matrix $\bm{C}$ in (\ref{eq-model-18}) defined by
\begin{equation}\label{eq-bound-3}
    \bm{C} = 
    \begin{bmatrix}
        \bm{W}_1^{\frac{1}{2}} \underline{\bm{E}}_{\mathcal{G}}^T  & \bm{O} \\
        \bm{O} &  \bm{W}_2^{\frac{1}{2}} 
    \end{bmatrix}
\end{equation} 
$\mathcal{S}$ can be equivalently formulated as
\begin{equation}\label{eq-model-30}
    \mathcal{S}(T_f) = \xi(T_f, \mathcal{s}) \frac{1}{T_f} \int_0^{T_f} \Delta \bm{y}^T \Delta \bm{y} \text{d} t
\end{equation}

With the assumption that $\bm{A}$ is Hurwitz and $\Delta \bm{x}(0) \!\!=\!\! \bm{0}$, for unit impulse disturbance inputs and white noise disturbance inputs with the unit covariance matrix, $\mathcal{S}(T_f)$ with $T_f = +\infty$ or the expectation of $\mathcal{S}(T_f)$ with $T_f \to +\infty$ is further equivalent to the square of the $\mathcal{H}_2$ norm of $\bm{G}(s)$. Mathematically, we have
\begin{equation}\label{eq-model-31}
    \mathcal{S}(+\infty) = \Vert \bm{G} \Vert_{\mathcal{H}_2}^2 ~\text{if}~ \Delta \bm{u}_i = \delta(t), \forall i \in \underline{\mathcal{V}}
\end{equation}
\begin{equation}\label{eq-model-32}
    \mathbb{E}\big[\! \lim_{T_f \to +\infty} \!\!\!\!  \mathcal{S}(T_f) \big] \!\!=\!\! \Vert \bm{G} \Vert_{\mathcal{H}_2}^2 
    \begin{aligned}
        ~ & \text{if}~ \mathbb{E}[ \Delta \bm{u}_i ] \!=\! 0, \forall i \in \underline{\mathcal{V}} ~\text{and}~ \\
        ~ & \mathbb{E}[ \Delta \bm{u}(t)  \Delta \bm{u}(t \!+\! \tau)^T  ] \!\!=\!\! \bm{I} \delta(\tau)
    \end{aligned}
\end{equation}
In (\ref{eq-model-31}), $\mathcal{s}$ is finite since $\Delta \bm{y}$ is bounded and $\lim_{t \to T_f} \Delta \bm{y}(t) = \Delta \bm{y} (0) = \bm{0}$, and thus $\xi(T_f, \mathcal{s}) = T_f$. In (\ref{eq-model-32}), $\xi(T_f, \mathcal{s}) = 1$ since $\mathcal{s}$ is infinite.

The $\mathcal{H}_2$ norm can be computed with observability Gramian $\bm{P}$ as
\begin{equation}\label{eq-model-33}
    \Vert \bm{G} \Vert_{\mathcal{H}_2}^2 = \text{Tr} (\bm{B}^T \bm{P} \bm{B} ) 
\end{equation}
Here $\bm{P}$ can be given by the following Lyapunov equation \citep{4-269}
\begin{equation}\label{eq-model-34}
    \bm{A}^T \bm{P} + \bm{P} \bm{A} + \bm{C}^T \bm{C} = \bm{O}
\end{equation}
Thereby $\Vert \bm{G} \Vert_{\mathcal{H}_2}$ can be interpreted as an integrated metric involving synchronization performance under different forms of disturbances, which will be employed for transmission switching.

\section{Impact of network structure on $\Vert \bm{G} \Vert_{\mathcal{H}_2}$}
To develope the approach of transmission switching to improve synchronization performance, in this section, we investigate how network structure impacts the synchronization performance metric. Specifically, bounds of $\Vert \bm{G} \Vert_{\mathcal{H}_2}$ are established for general cases and the close form under certain assumptions.

\subsection{Bounds of the synchronization performance metric}

\begin{thm}\label{theorem-1}
    (Laplacian-based bounds)
    Consider the system $(\bm{A}, \bm{B}, \bm{C})$ with $\bm{A}$, $\bm{B}$ and $\bm{C}$ given by (\ref{eq-bound-1}) to (\ref{eq-bound-3}) respectively, and $\bm{D}_{\rm{L}}$, $\underline{\bm{M}}_{\rm{SF}}$ and $\underline{\bm{D}}_{\rm{SF}}$ being all positive definite, and then $\Vert \bm{G} \Vert_{\mathcal{H}_2}$ satisfies
    \begin{equation}\label{theorem-1-1}
        \begin{aligned}
            & \frac{ \underline{\lambda}_{\rm{d}} }{2}
            \left[
                \Tr( \bm{\Pi} )
                + \Tr(\underline{\bm{M}}_{\rm{SF}}^{-1} \bm{W}_{2})
            \right]
            \leq 
            \Vert \bm{G} \Vert_{\mathcal{H}_2}^2 
            \\
            & \leq 
            \frac{\overline{\lambda}_{\rm{d}}}{2}
            \left[
                \Tr( \bm{\Pi} )
                + \Tr(\underline{\bm{M}}_{\rm{SF}}^{-1} \bm{W}_{2})
            \right]
        \end{aligned}
    \end{equation}
    where $ \bm{\Pi} =  \underline{\bm{L}}(\mathcal{G}, \bm{W}_1) \underline{\bm{L}}(\mathcal{G}, \bm{W}_{\rm{p}})^{-1} $, $\underline{\lambda}_{\rm{d}} = \min_{i \in \underline{\mathcal{V}}} \{ \frac{\Lambda_i}{d_i} \} $ and $\overline{\lambda}_{\rm{d}} = \max_{i \in \underline{\mathcal{V}}} \{ \frac{\Lambda_i}{d_i} \} $.
\end{thm}

\begin{pf}
    Partitioning $\bm{P}$ as
    \begin{equation}\label{proof-1-1}
        \bm{P} = 
        \begin{bmatrix}
            \bm{P}_{11}   & \bm{P}_{12} \\
            \bm{P}_{12}^T & \bm{P}_{22}
        \end{bmatrix}, 
    \end{equation}
    we have
    \begin{equation}\label{proof-1-2}
        \begin{aligned}
        \Vert \bm{G} \Vert_{\mathcal{H}_2}^2 
        \!=\! \Tr(  \bm{\Lambda}_{\rm{L}} \bm{D}_{\rm{L}}^{-2} \bm{T}_{\rm{L}}^T \bm{P}_{11} \bm{T}_{\rm{L}} )
            \!+\! \Tr( \bm{\Lambda}_{\rm{SF}} \underline{\bm{M}}_{\rm{SF}}^{-2} \bm{P}_{22} ) 
        % \\ 
        % & = \sum_{i \in \mathcal{V}_{\rm{L}}} \frac{\Lambda_i \bm{P}_{11, i} }{d_i^2} 
        %     +  \sum_{i \in \underline{\mathcal{V}}_{\rm{SF}}} \frac{\Lambda_i \bm{P}_{22, i} }{m_i^2} 
        \end{aligned}
    \end{equation}
    
    The Lyapunov equation (\ref{eq-model-34}) can be expanded as
    \begin{equation}\label{proof-1-3}
        \begin{aligned}
        \bm{A}^T \!\!
        \begin{bmatrix}
            \bm{P}_{11}   & \bm{P}_{12} \\
            \bm{P}_{12}^T & \bm{P}_{22}
        \end{bmatrix}
            \!+\! \begin{bmatrix}
                \bm{P}_{11}   & \bm{P}_{12} \\
                \bm{P}_{12}^T & \bm{P}_{22}
            \end{bmatrix}
            \bm{A} 
         \!+\! 
         \begin{bmatrix}
            \underline{\bm{L}}(\mathcal{G}, \bm{W}_1)  & \bm{O} \\
            \bm{O} &  \bm{W}_2 
        \end{bmatrix}
           \!=\! \bm{O}
        \end{aligned}
    \end{equation}
    
    Assume that the graph $\widetilde{\mathcal{G}}$ is connected and thus matrix $\underline{\bm{L}}(\mathcal{G}, \bm{W}_{\rm{p}})$ is nonsingular. Note that $\underline{\bm{L}}(\mathcal{G}, \bm{W}_{\rm{p}})$ is also symmetric. Right-multiplying equation (1,1) of (\ref{proof-1-3}) by $\underline{\bm{L}}(\mathcal{G}, \bm{W}_{\rm{p}})^{-1}$ gives
    \begin{equation}\label{proof-1-4}
        \begin{aligned}
            & 
            \underline{\bm{L}}(\mathcal{G}, \bm{W}_{\rm{p}})^T \bm{T}_{\rm{L}} \bm{D}_{\rm{L}}^{-1} \bm{T}_{\rm{L}}^T \bm{P}_{11}   \underline{\bm{L}}(\mathcal{G}, \bm{W}_{\rm{p}})^{-1} \\
            & + \underline{\bm{L}}(\mathcal{G}, \bm{W}_{\rm{p}})^T \bm{T}_{\rm{SF}} \underline{\bm{M}}_{\rm{SF}}^{-1} \bm{P}_{12}^T  \underline{\bm{L}}(\mathcal{G}, \bm{W}_{\rm{p}})^{-1} 
            \\
            &
            + \bm{P}_{11} \bm{T}_{\rm{L}} \bm{D}_{\rm{L}}^{-1} \bm{T}_{\rm{L}}^T \underline{\bm{L}}(\mathcal{G}, \bm{W}_{\rm{p}})   \underline{\bm{L}}(\mathcal{G}, \bm{W}_{\rm{p}})^{-1} \\
            & + \bm{P}_{12} \underline{\bm{M}}_{\rm{SF}}^{-1} \bm{T}_{\rm{SF}}^T \underline{\bm{L}}(\mathcal{G}, \bm{W}_{\rm{p}}) \underline{\bm{L}}(\mathcal{G}, \bm{W}_{\rm{p}})^{-1}
             = \bm{\Pi}
        \end{aligned}     
    \end{equation}
    
    With 1) the cyclic property of trace, 2) trace invariance of transpose, and 3) the equality that $\underline{\bm{L}}(\mathcal{G}, \bm{W}_{\rm{p}})   \underline{\bm{L}}(\mathcal{G}, \bm{W}_{\rm{p}})^{-1} = \bm{I}$, we obtain the following trace equality from (\ref{proof-1-4})
    \begin{equation}\label{proof-1-5}
      \Tr( \bm{T}_{\rm{L}} \bm{D}_{\rm{L}}^{-1} \bm{T}_{\rm{L}}^T \bm{P}_{11} ) 
       + \Tr ( \bm{P}_{12} \underline{\bm{M}}_{\rm{SF}}^{-1} \bm{T}_{\rm{SF}}^T ) 
       = \frac{1}{2} \Tr( \bm{\Pi} )
    \end{equation}
       
    Left-multiplying equation (2,2) of (\ref{proof-1-3}) by $\underline{\bm{M}}_{\rm{SF}}^{-1}$ gives that
    \begin{equation}\label{proof-1-8}
        \Tr(\underline{\bm{M}}_{\rm{SF}}^{-2} \underline{\bm{D}}_{\rm{SF}} \bm{P}_{22}) \!-\!\! \Tr( \underline{\bm{M}}_{\rm{SF}}^{-1} \bm{T}_{\rm{SF}}^T \bm{P}_{12} )  \!\!=\!\! \frac{1}{2}\! \Tr(\underline{\bm{M}}_{\rm{SF}}^{-1} \bm{W}_{2} \!)
    \end{equation}
    
    Since the matrices in a trace of a product can be switched without changing the result, we have $\Tr( \bm{P}_{12} \underline{\bm{M}}_{\rm{SF}}^{-1} \bm{T}_{\rm{SF}}^T ) = \Tr( \underline{\bm{M}}_{\rm{SF}}^{-1} \bm{T}_{\rm{SF}}^T \bm{P}_{12} )$. Thus 
    combining (\ref{proof-1-5}) and (\ref{proof-1-8}) gives that
    \begin{equation}\label{proof-1-9}
        \begin{aligned}
            & \Tr( \bm{T}_{\rm{L}} \bm{D}_{\rm{L}}^{-1} \bm{T}_{\rm{L}}^T \bm{P}_{11} ) 
            + \Tr(\underline{\bm{M}}_{\rm{SF}}^{-2} \underline{\bm{D}}_{\rm{SF}} \bm{P}_{22})\\
            & =  \frac{1}{2} \Tr( \bm{\Pi} ) + \frac{1}{2} \Tr(\underline{\bm{M}}_{\rm{SF}}^{-1} \bm{W}_{2})
        \end{aligned}
    \end{equation}

    Positive semi-definiteness of $\bm{P}$, $\bm{D}_{\rm{L}}$, $\underline{\bm{M}}_{\rm{SF}}$ and $\underline{\bm{D}}_{\rm{SF}}$ indicates that $\bm{D}_{\rm{L}} \bm{T}_{\rm{L}}^T \bm{P}_{11} \bm{T}_{\rm{L}}$ and $\underline{\bm{M}}_{\rm{SF}}^{-2} \underline{\bm{D}}_{\rm{SF}} \bm{P}_{22}$ are also positive semi-definite. Therefore by \citet[Theorem~2]{4-675}, the following inequality can be established from (\ref{proof-1-2}):
    \begin{equation}\label{proof-1-10}
        \! \begin{aligned}
            & \underline{\lambda}_{\rm{d, L}} \!\! \Tr(\!  \bm{D}_{\rm{L}}^{-1} \! \bm{T}_{\rm{L}}^T \! \bm{P}_{11}\! \bm{T}_{\rm{L}} )
            \!\!+\!\!  \underline{\lambda}_{\rm{d, SF}} \!\! \Tr( \underline{\bm{M}}_{\rm{SF}}^{-2} \underline{\bm{D}}_{\rm{SF}} \! \bm{P}_{22} ) 
            \!\!\leq\!\! 
            \Vert \bm{G} \Vert_{\mathcal{H}_2}^2 
            \\
            & \!\leq\! 
            \overline{\lambda}_{\rm{d, L}} \Tr(  \bm{D}_{\rm{L}}^{-1} \bm{T}_{\rm{L}}^T \bm{P}_{11} \bm{T}_{\rm{L}} )
            \!\!+\!\! \overline{\lambda}_{\rm{d, SF}}
             \Tr( \underline{\bm{M}}_{\rm{SF}}^{-2} \underline{\bm{D}}_{\rm{SF}} \bm{P}_{22} ) 
        \end{aligned}\!\!\!\!\!
    \end{equation}
    where $\underline{\lambda}_{\rm{d, L}} = \min_{i \in \mathcal{V}_{\rm{L}}} \{ \frac{\Lambda_i}{d_i} \} $ and $\overline{\lambda}_{\rm{d, L}} = \max_{i \in \mathcal{V}_{\rm{L}}} \{ \frac{\Lambda_i}{d_i} \} $, $\underline{\lambda}_{\rm{d, SF}} = \min_{i \in \underline{\mathcal{V}}_{\rm{SF}}} \{ \frac{\Lambda_i}{d_i} \} $ and $\overline{\lambda}_{\rm{d, SF}} = \max_{i \in \underline{\mathcal{V}}_{\rm{SF}}} \{ \frac{\Lambda_i}{d_i} \} $, 

    Furthermore, by relaxing $\underline{\lambda}_{\rm{d, L}}$ and $\underline{\lambda}_{\rm{d, SF}}$ to $\underline{\lambda}_{\rm{d}}$, and $\overline{\lambda}_{\rm{d, L}}$ and $\overline{\lambda}_{\rm{d, SF}}$ to $\overline{\lambda}_{\rm{d}}$, we have
    \begin{equation}\label{proof-1-11}
        \begin{aligned}
            & \underline{\lambda}_{\rm{d}} \left[ \Tr(\!  \bm{D}_{\rm{L}}^{-1} \! \bm{T}_{\rm{L}}^T \! \bm{P}_{11}\! \bm{T}_{\rm{L}} )
            \!\!+\!\!  \Tr( \underline{\bm{M}}_{\rm{SF}}^{-2} \underline{\bm{D}}_{\rm{SF}} \! \bm{P}_{22} ) \right]
            \!\!\leq\!\! 
            \Vert \bm{G} \Vert_{\mathcal{H}_2}^2 
            \\
            & \!\leq\! 
            \overline{\lambda}_{\rm{d}} \left[ \Tr(  \bm{D}_{\rm{L}}^{-1} \bm{T}_{\rm{L}}^T \bm{P}_{11} \bm{T}_{\rm{L}} )
            \!\!+\!\!
             \Tr( \underline{\bm{M}}_{\rm{SF}}^{-2} \underline{\bm{D}}_{\rm{SF}} \bm{P}_{22} ) \right]
        \end{aligned}.
    \end{equation}
    With the equality $\Tr(  \bm{D}_{\rm{L}}^{-2} \bm{T}_{\rm{L}}^T \bm{P}_{11} \bm{T}_{\rm{L}} ) \!\!=\!\! \Tr( \bm{T}_{\rm{L}} \bm{D}_{\rm{L}}^{-2} \bm{T}_{\rm{L}}^T \bm{P}_{11} ) $, substituting (\ref{proof-1-9}) into (\ref{proof-1-11}) gives the bounds in Theorem \ref{theorem-1}. \qed \end{pf}

\begin{remark}\label{remark-1}
    Analogue bounds were also derived by \citet{4-999-208} using a network-reduced dynamic model, in which, however, the influence of original network structure and load dynamics can not be observed. Note that bounds in Theorem \ref{theorem-1} are tighter than that derived by \citet{4-999-208}. Theorem \ref{theorem-1} reveals that $\Vert \bm{G} \Vert_{\mathcal{H}_2}$ can be impacted by damping and inertia of generators/inverters, damping of load, disturbance strength, and network structure embodied in $\underline{\bm{L}}(\mathcal{G}, \bm{W}_1)$ and $\underline{\bm{L}}(\mathcal{G}, \bm{W}_{\rm{p}})$. Corresponding to the first two factors, countermeasures including load-side control and allocating virtual inertia and damping have already been proved to be effective to enhance synchronism of low-inertia power grids. Transmission switching that changes network structure can promisingly achieve the same effect.
\end{remark}

In Theorem \ref{theorem-1}, bounds of $\Vert \bm{G} \Vert_{\mathcal{H}_2}$ are defined in terms of reduced Laplacian matrices of graph $\mathcal{G}$ that corresponds to the augmented power grid. To further state the bounds in terms of Laplacian matrices of graph $\widetilde{\mathcal{G}}$ that correspond to the unaugmented power grid, we partition $ \underline{\bm{L}}(\mathcal{G}, \bm{W}_{\rm{p}})$ as 
\begin{equation}\label{proof-2-1}
    \underline{\bm{L}}(\mathcal{G}, \bm{W}_{\rm{p}}) =
    \begin{bmatrix}
       \bm{L}_{\rm{HH}} & \bm{L}_{\rm{EH}}^T \\
       \bm{L}_{\rm{EH}} & \bm{L}_{\rm{EE}} 
    \end{bmatrix}
\end{equation}
where $ \bm{L}_{\rm{HH}} \!\!=\!\! \bm{T}_{\rm{SF}}^T \underline{\bm{L}}( \mathcal{G}_1, \bm{W}_{\rm{p1}} ) \bm{T}_{\rm{SH}} $, $ \bm{L}_{\rm{EH}} \!\!=\!\! \bm{T}_{\rm{L}}^T \underline{\bm{L}}( \mathcal{G}_1, \bm{W}_{\rm{p1}} ) \bm{T}_{\rm{SF}}$, and $ \bm{L}_{\rm{EE}} = \bm{L}(\widetilde{\mathcal{G}}, \widetilde{\bm{W}}_p ) + \bm{T}_{\rm{L}}^T \underline{\bm{L}}( \mathcal{G}_1, \bm{W}_{\rm{p1}} ) \bm{T}_{\rm{L}}$;  $\mathcal{G}_1 = \mathcal{G}_1(\mathcal{V}, \mathcal{E}_{\rm{SF}})$; $\bm{W}_{\rm{p1}} =  {\bm{B}}_{\rm{V},1} \frac{\partial \sin(\underline{\bm{E}}_{\mathcal{G}1}^{T} \bm{\alpha}^0) }{\partial (\underline{\bm{E}}_{\mathcal{G}1}^{T} \bm{\alpha}^0)  }  $ with $\underline{\bm{E}}_{\mathcal{G}1}$ formed by deleting the first row of the incidence matrix of $\mathcal{G}_1$ and $\bm{B}_{\rm{V}, 1} = \diag(V_i V_j b_{ij}), \forall (i,j) \!\in\! \mathcal{E}_{\rm{SF}}$.

Correspondingly, $\underline{\bm{L}}(\mathcal{G}, \bm{W}_1)$ can be formulated as the following block form
\begin{equation}\label{proof-2-4}
    \underline{\bm{L}}(\mathcal{G}, \bm{W}_1) =
    \begin{bmatrix}
       \bm{L}_{\rm{HH}}^* & \bm{L}_{\rm{EH}}^{*T} \\
       \bm{L}_{\rm{EH}}^* & \bm{L}_{\rm{EE}}^* 
    \end{bmatrix}.
\end{equation}

\begin{lem}\label{lemma-1}
    The following equality holds
    \begin{equation}\label{eq-lemma-1}
        \Tr( \bm{\Pi} ) = 
            \Tr( \bm{L}_{\rm{S}}^* \bm{L}_{\rm{S}}^{-1})
            +  \Tr(\bm{L}_{\rm{HH}}^* \bm{L}_{\rm{HH}}^{-1}) 
    \end{equation} 
    where $\bm{L}_{\rm{S}}^* = \bm{L}(\widetilde{\mathcal{G}}, \widetilde{\bm{W}}_1 ) + \bm{\Theta}^* $ and $\bm{L}_{\rm{S}} \!\!=\!\! \bm{L}(\widetilde{\mathcal{G}}, \widetilde{\bm{W}}_{\rm{p}} ) + \bm{\Theta}$; $\widetilde{\bm{W}}_1$ is the principal sub-matrix of $\bm{W}_1$ indexed by $\mathcal{E}_{\rm{L}}$; $\widetilde{\bm{W}}_{\rm{p}} = \widetilde{\bm{B}}_{\rm{V}} \frac{\partial \sin(\bm{E}_{\tilde{\mathcal{G}}}^{T} \bm{\alpha}_{\rm{L}}^0 ) }{\partial (\bm{E}_{\tilde{\mathcal{G}}}^{T} \bm{\alpha}_{\rm{L}}^0 )  }$ with $\widetilde{\bm{B}}_{\rm{V}} \!=\! \diag(V_i V_j b_{ij})$, $\forall (i,j) \!\in\! \mathcal{E}_{\rm{L}}$; $\bm{\Theta}^* \!\!=\!\! \diag( \bm{W}_{1, ij} , 0, ..., 0) \!\!\in\! \mathbb{R}^{|\mathcal{V}_{\rm{L}}| \times |\mathcal{V}_{\rm{L}}|}$ with bus $i$ being the angle reference bus and $j$ being its adjacent bus; $\bm{\Theta} \!\!=\!\! \diag( \bm{W}_{\rm{p1}}^{(1)} , 0, ..., 0) \!\!\in\! \mathbb{R}^{|\mathcal{V}_{\rm{L}}| \times |\mathcal{V}_{\rm{L}}|}$ with $\bm{W}_{\rm{p1}}^{(1)}$ being the first element of $\bm{W}_{\rm{p1}}$.
\end{lem}

\begin{pf}
    Under the assumption that $\Vert \diag(\bm{W}_{\rm{p1}}) \Vert_{-\infty} >0$, $\bm{L}_{\rm{HH}}$ is diagonal and also non-singular. Then the Schur complement of $\bm{L}_{\rm{HH}}$ is given by 
    \begin{equation}\label{proof-2-2}
        \begin{aligned}
            \bm{L}_{\rm{S}} 
            & = \bm{L}_{\rm{EE}} - \bm{L}_{\rm{EH}} \bm{L}_{\rm{HH}}^{-1} \bm{L}_{\rm{HE}}
            \\
            & = \bm{L}(\widetilde{\mathcal{G}}, \widetilde{\bm{W}}_p ) + \bm{T}_{\rm{L}}^T \underline{\bm{L}}( \mathcal{G}_1, \bm{W}_{\rm{p1}} ) \bm{T}_{\rm{L}} - \bm{L}_{\rm{EH}} \bm{L}_{\rm{HH}}^{-1} \bm{L}_{\rm{HE}}
            \\
            & = \bm{L}(\widetilde{\mathcal{G}}, \widetilde{\bm{W}}_p ) \!+\! \bm{T}_{\rm{L}}^T \underline{\bm{L}}( \mathcal{G}_1, \bm{W}_{\rm{p1}} ) \bm{T}_{\rm{L}} \!-\! \bm{E}_{\rm{I}} \bm{T}_{\rm{SF}}^T \underline{\bm{L}}( \mathcal{G}_1, \bm{W}_{\rm{p1}} ) \bm{T}_{\rm{L}}
            \\
            & = \bm{L}(\widetilde{\mathcal{G}}, \widetilde{\bm{W}}_p ) + \bm{\Theta}
        \end{aligned}
    \end{equation}
    where $\bm{E}_{\rm{I}} =  \bm{T}_{\rm{L}}^T \underline{\bm{E}}_{\mathcal{G}} \underline{\bm{E}}_{\mathcal{G}}^T  \bm{T}_{\rm{SF}} $.
       
    Furthermore, by \citet[Problem~3.7.11]{4-673} and equalities that $\bm{L}_{\rm{EH}} \bm{L}_{\rm{HH}}^{-1} = \bm{E}_{\rm{I}}$ and $\bm{L}_{\rm{HH}}^{-1} \bm{L}_{\rm{EH}}^T = \bm{E}_{\rm{I}}^T$, we have
    \begin{equation}\label{proof-2-3}
        \underline{\bm{L}}(\mathcal{G}, \bm{W}_{\rm{p}})^{-1} = 
        \begin{bmatrix}
            \bm{L}_{\rm{HH}}^{-1} +  \bm{E}_{\rm{I}}^T \bm{L}_{\rm{S}}^{-1} \bm{E}_{\rm{I}}
            & - \bm{E}_{\rm{I}}^T \bm{L}_{\rm{S}}^{-1}
            \\
            - \bm{L}_{\rm{S}}^{-1} \bm{E}_{\rm{I}}
            & \bm{L}_{\rm{S}}^{-1}
        \end{bmatrix}
    \end{equation}
    Substituting (\ref{proof-2-3}) and (\ref{proof-2-4}) into $\bm{\Pi}$ yields
    \begin{equation}\label{proof-2-5}
        \begin{aligned}
             \Tr(\bm{\Pi}) = &  \Tr( (\bm{L}_{\rm{EE}}^* - \bm{L}_{\rm{EH}}^*  \bm{E}_{\rm{I}}^T) \bm{L}_{\rm{S}}^{-1}) +  \Tr(\bm{L}_{\rm{HH}}^* \bm{L}_{\rm{HH}}^{-1}) 
            \\
            &  + \Tr( (\bm{L}_{\rm{HH}}^* \bm{E}_{\rm{I}}^T - \bm{L}_{\rm{HE}}^*) \bm{L}_{\rm{S}}^{-1} \bm{E}_{\rm{I}}) 
        \end{aligned}
    \end{equation}
    which together with $ \bm{L}_{\rm{S}}^* = \bm{L}_{\rm{EE}}^* - \bm{L}_{\rm{EH}}^*  \bm{E}_{\rm{I}}^T$ and $\bm{L}_{\rm{HH}}^* \bm{E}_{\rm{I}}^T = \bm{L}_{\rm{HE}}^*$ gives Lemma \ref{lemma-1}. \qed
\end{pf}

\begin{remark}
    Matrix $\bm{L}_{\rm{S}}$ can be interpreted as the Laplacian matrix of graph $\widetilde{\mathcal{G}}$ added one self-loop at the load node connected with the angle reference bus. The weight of the self-loop equals to the nonzero elements in $\bm{\Theta}$ which is positive. Matrix $\bm{L}_{\rm{S}}^*$ is analogue.
\end{remark}

By Lemma \ref{lemma-1}, bounds in Theorem \ref{theorem-1} is restate in terms of Laplacian matrices of $\widetilde{\mathcal{G}}$ as follows:

\begin{cor}\label{corollary-1}
    Consider the system $(\bm{A}, \bm{B}, \bm{C})$ with $\bm{A}$, $\bm{B}$ and $\bm{C}$ given by (\ref{eq-bound-1}) to (\ref{eq-bound-3}) respectively, and $\bm{D}_{\rm{L}}$, $\underline{\bm{M}}_{\rm{SF}}$ and $\underline{\bm{D}}_{\rm{SF}}$ being all positive definite, and then $\Vert \bm{G} \Vert_{\mathcal{H}_2}$ satisfies
    \begin{equation}\label{corollary-1-1}
        \begin{aligned}
            & \frac{ \underline{\lambda}_{\rm{d}} }{2}
            \left[
                \Tr( \bm{L}_{\rm{S}}^* \bm{L}_{\rm{S}}^{-1})
                \!+\!\!  \Tr(\bm{L}_{\rm{HH}}^* \bm{L}_{\rm{HH}}^{-1} \!+\! \underline{\bm{M}}_{\rm{SF}}^{-1} \bm{W}_{2})
            \right]
            \!\! \leq \!\!
            \Vert \bm{G} \Vert_{\mathcal{H}_2}^2 
            \\
            & \leq 
            \frac{\overline{\lambda}_{\rm{d}}}{2}
            \left[
                \Tr( \bm{L}_{\rm{S}}^* \bm{L}_{\rm{S}}^{-1})
                +  \Tr(\bm{L}_{\rm{HH}}^* \bm{L}_{\rm{HH}}^{-1} + \underline{\bm{M}}_{\rm{SF}}^{-1} \bm{W}_{2})
            \right]
        \end{aligned}
    \end{equation}
\end{cor}

\begin{remark}
    In each bound in Corollary \ref{corollary-1}, only the first trace term is dependent on structure of the unaugmented power girds. $\bm{L}_{\rm{S}}^*$ and $\bm{L}_{\rm{S}}$ are Laplacian matrices of $\widetilde{\mathcal{G}}$ add one self-loop, whose edge weights are related to weighting factors in $\mathcal{S}$ and active power flow at the equilibrium point, respectively.
\end{remark}

Furthermore, under Assumption \ref{assumption-1}, the synchronization performance metric $\Vert \bm{G} \Vert_{\mathcal{H}_2}$ can be formulated in close form, for which the observability Gramian is eliminated.
\begin{assumption}\label{assumption-1}
    The ratio of disturbance strength to load damping and that of disturbance strength to generator/inverter damping are uniform, i.e., $\frac{\Lambda_i}{d_i} \!\!=\!\! \lambda_{\rm{d}}, \forall i \!\!\in\!\! \mathcal{V}_{\rm{L}} \cup \underline{\mathcal{V}}_{\rm{SF}}$.
\end{assumption}

\begin{cor}\label{corollary-2}
    ($\Vert \bm{G} \Vert_{\mathcal{H}_2}$ under Assumption \ref{assumption-1}) 
    Consider the system $(\bm{A}, \bm{B}, \bm{C})$ with $\bm{A}$, $\bm{B}$ and $\bm{C}$ given by (\ref{eq-bound-1}) to (\ref{eq-bound-3}) respectively, Assumption \ref{assumption-1} satisfied, and $\bm{D}_{\rm{L}}$, $\underline{\bm{M}}_{\rm{SF}}$ and $\underline{\bm{D}}_{\rm{SF}}$ being all positive definite. Then
    \begin{equation}\label{corollary-2-1}
        \Vert \bm{G} \Vert_{\mathcal{H}_2}^2 
        \!\!=\!\! 
        \frac{\lambda_{\rm{d}} }{2}
        \left[
            \Tr( \bm{L}_{\rm{S}}^* \bm{L}_{\rm{S}}^{-1})
            \!\!+\!\!  \Tr(\bm{L}_{\rm{HH}}^* \bm{L}_{\rm{HH}}^{-1} \!+\! \underline{\bm{M}}_{\rm{SF}}^{-1} \bm{W}_{2})
        \right]
    \end{equation}
\end{cor}

\begin{pf}
    Assumption \ref{assumption-1} indicates that $ \underline{\lambda}_{\rm{d}} =  \overline{\lambda}_{\rm{d}} =  \lambda_{\rm{d}}$. Then by Corollary \ref{corollary-1}, we conclude Corollary \ref{corollary-2}. \qed
\end{pf}

\begin{remark}
    For practical power grids, Assumption \ref{assumption-1} is reasonable since that strength of disturbances is approximately proportional to the load power or generation power while the same for the damping of loads and generators/inverters. Or to be exact, Theorem \ref{theorem-1} and Corollary \ref{corollary-2} provide tight bounds of $\Vert \bm{G} \Vert_{\mathcal{H}_2}$ for practical power grids since $\underline{\lambda}_{\rm{d}}\approx \overline{\lambda}_{\rm{d}}$.
\end{remark}

\section{Sensitivity-based transmission switching approach}
The transmission switching problem is traditionally tackled by optimization-based approaches where some steady-state metrics are generally concerned. However, finding the optimal network topology that minimizes $\Vert \bm{G} \Vert_{\mathcal{H}_2}$ is far from easy whether based on the Lyapunov equation or the close-form $\Vert \bm{G} \Vert_{\mathcal{H}_2}$. In this section, we develop a transmission switching approach by analyzing the sensitivity of $\Vert  \bm{G} \Vert_{\mathcal{H}_2}^2$ or its bounds to perturbation of network susceptance.

\subsection{Sensitivity Analysis}
Since the sensitivity of close-form $\Vert  \bm{G} \Vert_{\mathcal{H}_2}^2$ its proportional to that of its bounds, we only focus on the former. Sensitivity of $\Vert  \bm{G} \Vert_{\mathcal{H}_2}^2$ to perturbation of branch susceptance is given by the following proposition.
\begin{prop}\label{prop-1}
    (${\partial \Vert  \bm{G} \Vert_{\mathcal{H}_2}^2 }/{ \partial b_{ij} } $ under Assumption \ref{assumption-1})
    Consider the system $(\bm{A}, \bm{B}, \bm{C})$ with $\bm{A}$, $\bm{B}$ and $\bm{C}$ given by (\ref{eq-bound-1}) to (\ref{eq-bound-3}) respectively, Assumption \ref{assumption-1} satisfied, and $\bm{D}_{\rm{L}}$, $\underline{\bm{M}}_{\rm{SF}}$ and $\underline{\bm{D}}_{\rm{SF}}$ being all positive definite. Then $\forall (i,j) \in \mathcal{E}_{\rm{L}}$, we have
    \begin{equation}\label{eq-prop-1}
        \frac{\partial \Vert  \bm{G} \Vert_{\mathcal{H}_2}^2 }{ \partial b_{ij} } = 
        - \frac{\lambda_{\rm{d}} }{2} \Tr( \bm{L}_{\rm{S}}^{-1} \bm{L}_{\rm{S}}^* \bm{L}_{\rm{S}}^{-1}  \bm{L}(\widetilde{\mathcal{G}}, \widetilde{\bm{W}}_p \bm{E}^{ij} ) )
        < 0
    \end{equation}
    where $\bm{E}^{ij} \in \mathbb{R}^{|\mathcal{E}_{\rm{L}}| \times |\mathcal{E}_{\rm{L}}|}$ is a diagonal matrix with only one non-zero element being $\frac{1}{b_{ij}}$ for edge $(i,j)$.
\end{prop}

\begin{pf}
    In the right-hand side of (\ref{corollary-2-1}), network susceptance only appears in $\bm{L}_{\rm{S}}$. Thus
\begin{equation}\label{eq-sens-1}
    \frac{\partial \Vert  \bm{G} \Vert_{\mathcal{H}_2}^2 }{ \partial b_{ij} } = \frac{\lambda_{\rm{d}} }{2} \frac{ \partial \Tr( \bm{L}_{\rm{S}}^* \bm{L}_{\rm{S}}^{-1}) }{\partial b_{ij} } = 
    \frac{\lambda_{\rm{d}} }{2} \Tr( \bm{L}_{\rm{S}}^* \frac{ \partial \bm{L}_{\rm{S}}^{-1} }{\partial b_{ij} } )
\end{equation}
with
\begin{equation}\label{eq-sens-2}
    \frac{ \partial \bm{L}_{\rm{S}}^{-1} }{\partial b_{ij} } = - \bm{L}_{\rm{S}}^{-1}  \frac{\partial \bm{L}_{\rm{S}} }{\partial b_{ij} } \bm{L}_{\rm{S}}^{-1} = - \bm{L}_{\rm{S}}^{-1}  \bm{L}(\widetilde{\mathcal{G}}, \widetilde{\bm{W}}_p \bm{E}^{ij} ) \bm{L}_{\rm{S}}^{-1},
\end{equation}
which together with the cyclic property of trace gives the equality in (\ref{eq-prop-1}).

Furthermore, by \citet[Lemma~1]{4-676}, $\bm{L}_{\rm{S}} \!\!\succ\!\! 0$, $\bm{L}_{\rm{S}}^* \!\!\succ\!\! 0$ and thus $\bm{L}_{\rm{S}}^{-1} \!\!\succ\!\! 0$. Therefore, $\bm{L}_{\rm{S}}^{-1} \bm{L}_{\rm{S}}^* \bm{L}_{\rm{S}}^{-1} \!\!\succ\!\! 0$ and its smallest eigenvalue $\lambda_{\min} \!\!>\!\! 0$. $ \bm{L}(\widetilde{\mathcal{G}}, \widetilde{\bm{W}}_p \bm{E}^{ij} ) $ is with eigenvalues $\lambda_0 \!=\! \lambda_1 \!=\! \cdots \!=\! \lambda_{|\mathcal{V}_{\rm{L}}| - 1} \!=\! 0 \!<\! \lambda_{|\mathcal{V}_{\rm{L}}|}$, which gives that $\Tr( \bm{L}(\widetilde{\mathcal{G}}, \widetilde{\bm{W}}_p \bm{E}^{ij} ) )  > 0$. Then by \citet[Theorem~2]{4-675}, the following inequality holds:
\begin{equation}\label{eq-sens-3}
     \Tr(\! \bm{L}_{\rm{S}}^{-1} \! \bm{L}_{\rm{S}}^* \bm{L}_{\rm{S}}^{\!-1} \!\! \bm{L}(\!\widetilde{\mathcal{G}},\! \widetilde{\bm{W}}_{\!\!p} \bm{E}^{ij} ) ) \!\!\geq\!\! \lambda_{\min} \!\!\Tr(\! \bm{L}(\widetilde{\mathcal{G}},\! \widetilde{\bm{W}}_{\!\!p} \bm{E}^{ij} ) )  \!\!>\! 0 \!\!\!
\end{equation}
which together with $\lambda_{\rm{d}} > 0$ concludes the inequality in (\ref{eq-prop-1}).  \qed
\end{pf} 

\begin{remark}
    Proposition \ref{prop-1} indicates that under Assumption \ref{assumption-1}, a positive perturbation of branch susceptance always decreases $\Vert \bm{G} \Vert_{\mathcal{H}_2}$. 
\end{remark}

\subsection{Transmission Switching Approach}
The transmission switching problem considered to solve is described as follows:
\begin{problem}\label{prob-1}
    Given $\bm{p}_{\rm{in}}^0$, a initial network topology $\widetilde{\mathcal{G}}(\mathcal{V}, \mathcal{E}_{\rm{L, u}} ) $ where $\mathcal{E}_{\rm{L, u}}$ is the set of undispatchable branches, and a dispatchable line set $\mathcal{E}_{\rm{L, s}} $, find a line set $\mathcal{E}_{\rm{L, on}}\!\!\subset\!\! \mathcal{E}_{\rm{L, s}}$ satisfying $|\mathcal{E}_{\rm{L, on}}| \!\!=\!\! n_{\rm{on}}$ with $n_{\rm{on}}$ being the maximum number of lines to switch on, such that $\widetilde{\mathcal{G}}(\mathcal{V}\!, \mathcal{E}_{\rm{L, u}} \!\cup \mathcal{E}_{\rm{L, on}} ) $ minimizes $\Vert \bm{G} \Vert_{\mathcal{H}_2}$.
\end{problem}

In theory, Problem \ref{prob-1} is NP-hard while for practical implementation, it is preferred to obtain a good solution within the time available. Based on the sensitivity analysis, we can easily determine the next best line to switch on, which cannot guarantee optimality but may provide good solutions fast. Thereby we can develop an approach to solve Problem \ref{prob-1} by switching on one line at a time, which is given by Algorithm \ref{alg-1}. A total of $n_{\rm{on}}$ iterations are required, in each of which we mainly need to compute power flow and inverse of $\bm{L}_{\rm{S}}$ both once.

\begin{algorithm}[htb]
    \caption{Transmission switching approach.}
    \label{alg-1}
    \begin{algorithmic}[1]
      \Require
      $\bm{p}_{\rm{in}}^0$, $\widetilde{\mathcal{G}}(\mathcal{V}, \mathcal{E}_{\rm{L, u}} ) $, $\mathcal{E}_{\rm{L, s}} $, $n_{\rm{on}}$
        
      \Ensure
      $\mathcal{E}_{\rm{L, on}} $
        
      \State Initialize $\mathcal{E}_{\rm{L, on}} \gets \emptyset$;
      \Repeat
      \State  { $\widetilde{\bm{W}}_{\rm{p}}  \gets $  Compute power flow for $\widetilde{\mathcal{G}}(\mathcal{V}_{\rm{L}}, \mathcal{E}_{\rm{L, u}} \!\cup\! \mathcal{E}_{\rm{L, s}} ) $ {\color{white} 111}with $\forall (i,j) \in \mathcal{E}_{\rm{L, s}} \!-\! \mathcal{E}_{\rm{L, on}}$, $b_{ij} \!=\! 0$ and $\forall (i,j) \!\in\! \mathcal{E}_{\rm{L, on}}$, {\color{white} 1...} $b_{ij}$ being its actual value;} 
      \State Compute $\bm{L}_{\rm{S}}^{-1}$ and $\bm{L}_{\rm{S}}^*$;
      \State $\forall (i,j) \in \mathcal{E}_{\rm{L, s}} \!-\! \mathcal{E}_{\rm{L, on}} $, compute ${\partial \Vert  \bm{G} \Vert_{\mathcal{H}_2}^2 }/{ \partial b_{ij} } $;
      \State $\mathcal{E}_{\rm{L, on}} \gets \mathcal{E}_{\rm{L, on}} + \argmax_{(i,j) \in \mathcal{E}_{\rm{L, s}} \!-\! \mathcal{E}_{\rm{L, on}} } {\partial \Vert  \bm{G} \Vert_{\mathcal{H}_2}^2 }/{ \partial b_{ij} } $

      \Until $|\mathcal{E}_{\rm{L, on}}| = n_{\rm{on}}$ \\

      \Return $\mathcal{E}_{\rm{L, on}}$
    \end{algorithmic}
  \end{algorithm}

\section{Numerical Examples}

\begin{figure}
    \begin{center}
    \includegraphics[width=7.35cm, height = 4.75cm]{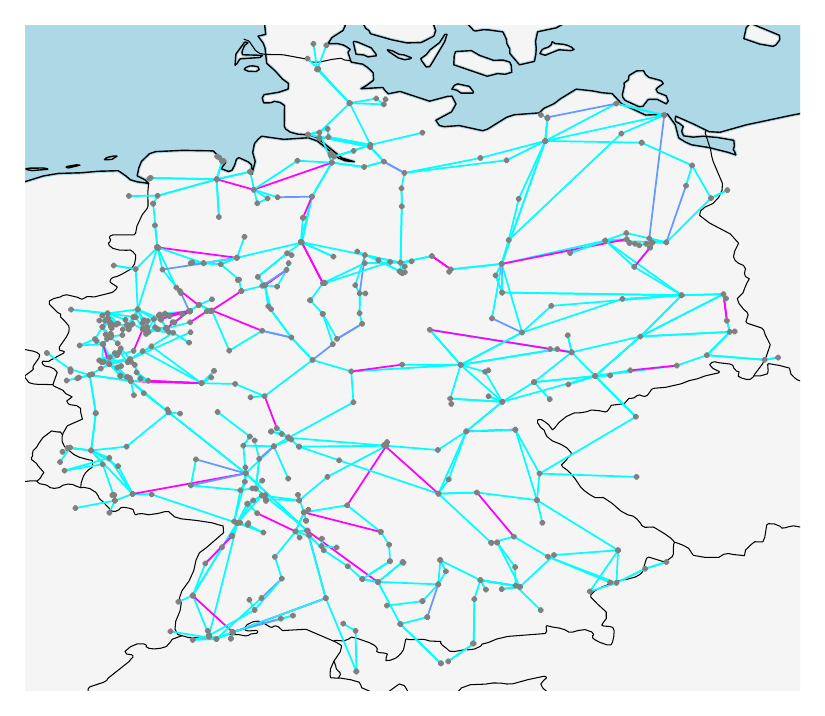}    % The printed column width is 8.4 cm.
    \caption{Topology of the SciGRID network for Germany.} 
    \label{fig-1}
    \end{center}
    \end{figure}

In the following, the transmission switching approach to improve synchronization of power grids is tested using the SciGRID network for Germany with its load snapshot at 12:00:00 January 1st, 2011 \citep{4-677} and dispatch of generators being optimized by the linear OPF. Fig.~\ref{fig-1} shows the main topology of the grid. This grid contains 585 buses including 489 generator/inverter buses and 485 load buses, 852 lines (multi-circuits lines are transformed into one-circuit lines) and 96 transformers. To carry out transmission switching, it is assumed that 60 lines (colored by purple and blue in Fig.~\ref{fig-1}) are dispatchable and $n_{\rm{on}} \!=\! 20$.

\subsection{Switching results}

\begin{figure}
    \begin{center}
    \includegraphics[width=7cm]{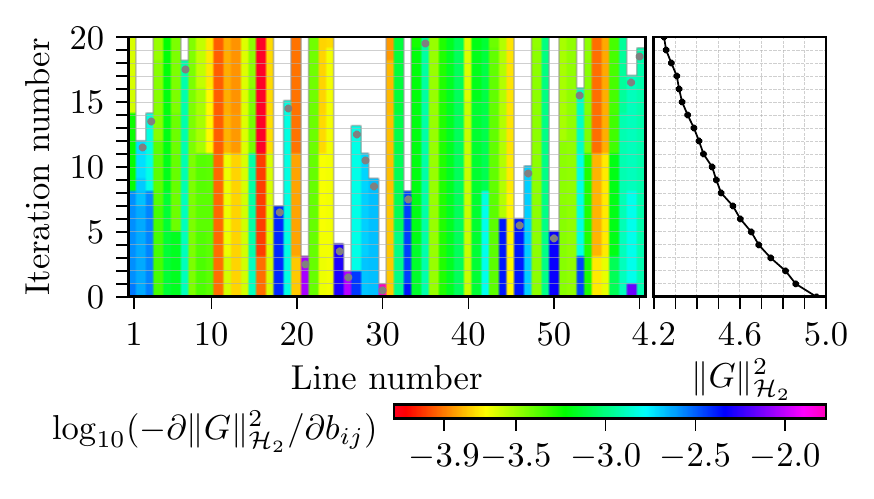}    % The printed column width is 8.4 cm.
    \caption{Sensitivity values in each iteration (left) and change of $\Vert \bm{G} \Vert_{\mathcal{H}_2}^2 $ with lines switched on in turn (right). Lines to switch on are marked by grey dots in the left figure.} 
    \label{fig-2}
    \end{center}
\end{figure}

Fig.~\ref{fig-2} (left) shows sensitivity values of dispatchable lines during iteration, and the line marked by a grey dot is with the largest sensitivity within the current iteration and selected to switch on. All lines to switch on, i.e., $\mathcal{E}_{\rm{L, on}}$, is also colored by blue in Fig.~\ref{fig-1}. It is found that sensitivity values of some lines (e.g., line 1 and line 59) vary widely during iteration, which indicates that switching of one line can influence the potential of remaining dispatchable lines to improve synchronization performance. This influence prevents us increasing the number of lines selected to switch on in each iteration, which though can accelerate computation. 

Fig.~\ref{fig-2} (right) is the change curve of $\Vert \bm{G} \Vert_{\mathcal{H}_2}^2$ with lines selected in each iteration switched on in turn. With more lines switched on, synchronization performance is continually improved while overall, the absolute value of slope of the curve decreases. These two trends correspond with negativeness of 
${\partial \Vert  \bm{G} \Vert_{\mathcal{H}_2}^2 }/{ \partial b_{ij} }$ and a observation from Fig.~\ref{fig-2} (left) that ${\partial \Vert  \bm{G} \Vert_{\mathcal{H}_2}^2 }/{ \partial b_{ij} }$ decreases overall as the iteration number increases, respectively. 

% Here rais the question that when shoud we used switcing as a way ...

\subsection{Output Response to time-varying disturbances}

We further compare output response of the power grid with and without lines in $\mathcal{E}_{\rm{L, on}}$ switched on. As mentioned in Section \ref{section-3}, $\Vert \bm{G} \Vert_{\mathcal{H}_2}^2$ implies synchronization performance in terms of white noise disturbance inputs which in fact, never disturb physical power grids. Thus taking into account the actual situation, $\forall i \!\!\in\!\! \underline{\mathcal{V}}$, we set $\Delta \bm{u}_i$ as a time-varying signal which changes its value randomly at a equal interval 2 s according to a truncated normal distribution with mean 0, variance 1 and interval $[-1, 1]$.
%are used as a theoretical approximation for power fluctuation

\begin{figure}
    \begin{center}
    \includegraphics[width=7.5cm]{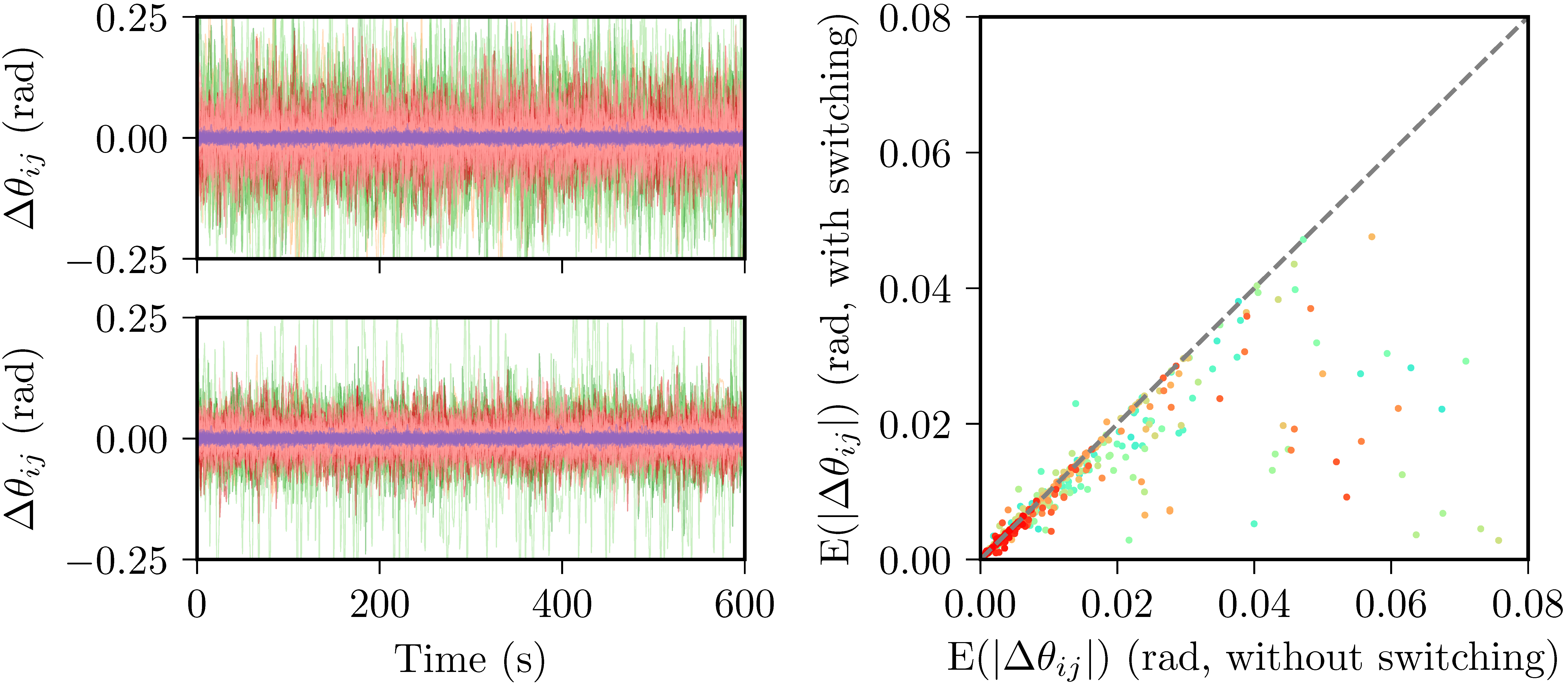}    % The printed column width is 8.4 cm.
    \caption{Output response $\Delta \theta_{ij} = \Delta \theta_i - \Delta \theta_j$ to time-varying disturbances with (left top) and without (left bottom) line switching, and scatter plot of $\rm{E}(|\Delta \theta_{ij}|)$ (right).} 
    \label{fig-3}
    \end{center}
\end{figure}

\begin{figure}
    \begin{center}
    \includegraphics[width=7.5cm]{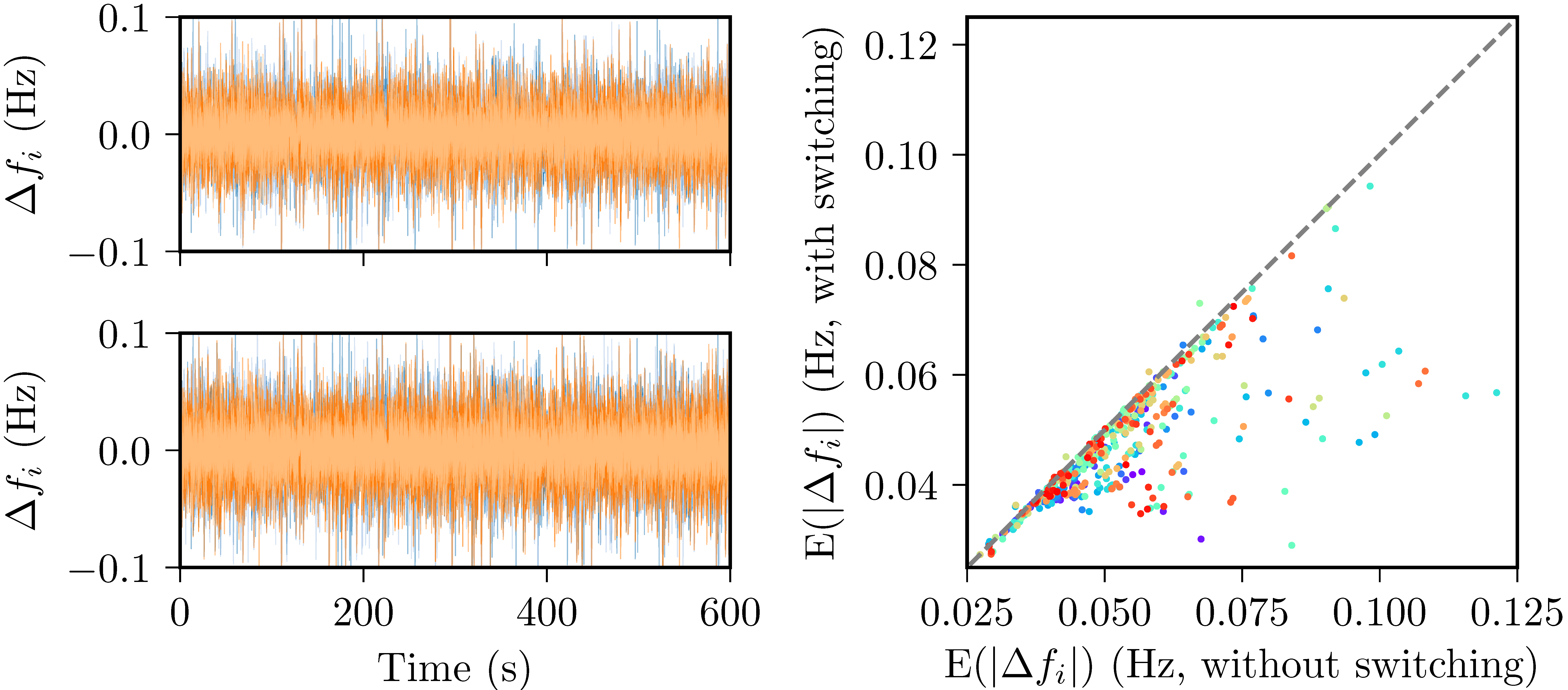}    % The printed column width is 8.4 cm.
    \caption{Output response $\Delta f_i \!\!=\!\! \frac{\Delta \omega}{ 2 \pi} $ to time-varying disturbances with (left top) and without (left bottom) line switching, and scatter plot of ${\rm{E}}(|\Delta f_i|)$ (right).} 
    \label{fig-4}
    \end{center}
\end{figure}

Fig.~\ref{fig-3} and Fig.~\ref{fig-4} show the output response to the time-varying disturbances with and without line switching. In the scatter plots, $\rm{E}(|\Delta \theta_{ij}|)$ (or ${\rm{E}}(|\Delta f_i|)$) is the mean of $\Delta \theta_{ij}$ ( or ${\rm{E}}(|\Delta f_i|)$) obtained by sampling the output response, and a dot under the dashed line indicates that corresponding $\rm{E}(|\Delta \theta_{ij}|)$ ( or ${\rm{E}}(|\Delta f_i|)$) is reduced after switching on lines in $\mathcal{E}_{\rm{L, on}}$. In Fig.~\ref{fig-3} (left), a distinct shrink of the curve cluster, connoting improvement of phase cohesiveness, can be observed after lines being switched on. Fig.~\ref{fig-4} (left) provides no obvious indication of change of frequency synchronization performance. Fig.~\ref{fig-3} (right) and Fig.~\ref{fig-4} (right) both show that most dots are below the dashed line and therefore, phase cohesiveness of most branches and frequency synchronization performance of most generators/inverters are both improved by line switching. Phase cohesiveness of several branches is slightly undermined after lines being switched on and in general, branches with the worst pre-switch phase cohesiveness present the greatest performance improvement. Frequency synchronization performance is analogous.

\section{Conclusion}

In response to new challenges caused by transition of power girds, structure-oriented control and optimization should play a more important role than ever before. In this paper, we propose to utilize transmission switching as a mean to improve synchronization performance of grids and develope a sensitivity-based switching approach. However, it should be pointed out that the transmission switching approach developed in this paper is still far from practical application. Impact of switching on other aspects of system performance, such as transient stability and line overload, should be considered while determining lines to switch. Switching approaches that are able to tackle more general switching scenarios are expected. Coordination of transmission switching and regulation of node dynamic properties could be the final package to pursue.

% A conclusion section is not required. Although a conclusion may review
% the main points of the paper, do not replicate the abstract as the
% conclusion. A conclusion might elaborate on the importance of the work
% or suggest applications and extensions. 

% \begin{ack}
% Place acknowledgments here.
% \end{ack}

\bibliography{References/4}             % bib file to produce the bibliography
                                                     % with bibtex (preferred)
                                                   
%\begin{thebibliography}{xx}  % you can also add the bibliography by hand

%\bibitem[Able(1956)]{Abl:56}
%B.C. Able.
%\newblock Nucleic acid content of microscope.
%\newblock \emph{Nature}, 135:\penalty0 7--9, 1956.

%\bibitem[Able et~al.(1954)Able, Tagg, and Rush]{AbTaRu:54}
%B.C. Able, R.A. Tagg, and M.~Rush.
%\newblock Enzyme-catalyzed cellular transanimations.
%\newblock In A.F. Round, editor, \emph{Advances in Enzymology}, volume~2, pages
%  125--247. Academic Press, New York, 3rd edition, 1954.

%\bibitem[Keohane(1958)]{Keo:58}
%\newblock \emph{Power and Interdependence: World Politics in Transitions}.
%\newblock Little, Brown \& Co., Boston, 1958.

%\bibitem[Powers(1985)]{Pow:85}
%\newblock Is there a way out?
%\newblock \emph{Harpers}, pages 35--47, June 1985.

%\bibitem[Soukhanov(1992)]{Heritage:92}
%\newblock \emph{{The American Heritage. Dictionary of the American Language}}.
%\newblock Houghton Mifflin Company, 1992.

%\end{thebibliography}

% \appendix
% \section{A summary of Latin grammar}    % Each appendix must have a short title.
% \section{Some Latin vocabulary}              % Sections and subsections are supported  
                                                                         % in the appendices.
\end{document}